# Multisoliton solutions via a separation of variables

B. P. Ryssev

**Abstract**
It is shown that multisoliton solutions of several well known nonlinear PDEs(x, t) can be obtained by certain separation of variables: each n-soliton arises from a mutual solution of a nonlinear ODE(x), common for all NPDEs considered, and linear PDE(x, t) with the linear operator from the NPDE in question.

The main results of this paper:
• A solution of the n-th nonlinear ordinary differential equation (NODE) from an infinite ordered set defines the functional form, the x-dependence, of the n-soliton for several nonlinear one-dimensional partial differential equations (NPDEs). Individual is only the time-dependence of such solution that comes from a linear PDE with the linear operator from a given NPDE (§1, 2).
• The NPDEs admitting such a separation of variables (SV) can be derived from the ground up. There are two types of them corresponding to two types of NODEs (§3).
• Each original NODE can be transformed into a NODE for an auxiliary function. The general ansatz that solves the latter has also a particular form, which gives rise to the special, easy to find for any n, multisoliton solutions (§4-6).
• At least the first few of NODE for an auxiliary function can be reduced to linear ODEs with constant coefficients (§7).
    The material of this paper is an extension and generalization of our approach proposed in [1].

**1.** The Korteweg-de Vries (KdV) and modified KdV (mKdV) equations are used as the first examples of the two types of NPDEs admitting such SV. We begin with one-soliton.
    Two nonlinear ODEs (m = 1, 2) for a function u(x) (the primes denote the x-derivatives)
$$H_1^{[m]} \equiv (\ln u)'' + u^m = 0$$
have the soliton-like solutions (their notation corresponds to that of $H_1^{[1,2]}$)
$$u_1^{[1]}(x) = 2k_1^2 \operatorname{sech}^2(k_1 x + x_1) \qquad u_1^{[2]}(x) = k_2 \operatorname{sech}(k_2 x + x_2) \qquad (1)$$
These $H_1^{[m]}$ appear in the left-hand sides $U^{[m]}$ of the KdV (m = 1) and mKdV (m = 2) for u(x, t)
$$U^{[m]} \equiv Qu + 6u^m u' \qquad Q = \partial_t + \partial^3 \qquad \partial_t = \partial/\partial t, \ \partial^i = \partial^i/\partial x^i \qquad (2)$$
as follows. Let r = 1/u, l = ln u, use rQu = Ql + 3l'l'' + l'³ to get rQu + uQr = 6l'l''. This, with Qu from (2), yields
$$rU^{[m]} + uR^{[m]} = 0 \qquad R^{[m]} \equiv Qr + 6H_1^{[m]} r' \qquad (3)$$
(simply put, $U^{[m]}/u^2 = -R^{[m]}$). Thus, u(x, t) is a solution of $U^{[m]} = 0$, if r(x, t) is that of $R^{[m]} = 0$. In the latter (for each m), we set separately
$$H_1^{[m]} = 0 \qquad Qr = 0 \qquad (4)$$
and seek for a mutual solution $r_1^{[m]}(x, t)$ of these two equations. That of the first follows from (1)
$$r_1^{[1]} = (1/4k_1^2)[\cosh 2(k_1 x + x_1) + 1] \qquad r_1^{[2]} = (1/k_2)\cosh(k_2 x + x_2)$$
Set $x_m = -\omega_m t$, $\omega_m = $ const, then $r_1^{[m]}$ satisfy the second if $2\omega_1 = (2k_1)^3$ and $\omega_2 = k_2^3$. With such $x_m$, the functions in (1) represent the well known one-soliton solutions of the KdV and mKdV equations
$$u_1^{[1]}(x, t) = 2k_1^2 \operatorname{sech}^2(k_1 x - 4k_1^3 t) \qquad u_1^{[2]}(x, t) = k_2 \operatorname{sech}(k_2 x - k_2^3 t) \qquad (5)$$

Thus, by construction, the nonlinear ODE determines the functional form (x-dependence) of each of them, whereas the linear PDE just defines $\omega_m$ involved in its t-dependence.

What is more, the nonlinear ODEs $H_1^{[m]} = 0$ can be reduced to the simplest linear ones for $r(x)$. For this, write them in terms of r: $-r^2 H_1^{[m]} \equiv r^2 (\ln r)'' - u^{m-2} = (rr'' - r'^2) - r^{2-m} = 0$. Apply $(1/r^2)\partial$ (differentiate and then divide by $r^2$) to get $(r''/r)' - (r^{2-m})'/r^2 = 0$. Integration, with constant $c_m$, yields the sought ODEs

$$m = 1:\ r'' + 1 = c_1 r \qquad\qquad m = 2:\ r'' = c_2 r \qquad\qquad (6)$$

They have, respectively, the solutions $b_1 \cosh(q_1 x + x_1) + b$ and $b_2 \cosh(q_2 x + x_2)$ with $q_m^2 = c_m$, $b = 1/c_1$ and arbitrary $b_m$. The latter can be found by substituting these solutions back into $H_1^{[m]} = 0$, which gives $b = (b_1 q_1)^2 = 1/q_1^2$ and $(b_2 q_2)^2 = 1$, cf. §5. Such solutions include the above $r_1^{[m]}$. (Here and below, we do not consider solutions containing sinh).

Conventions. The plural form mark "s", like in $U^{[m]}$s or $b_m$'s, is omitted when such a form is obvious from context; "lhs" and "rhs" stand for left and right-hand-side.

**2.** The multisolitons of the KdV and mKdV can be obtained in a similar manner. We introduce two ($m = 1, 2$) infinite sets ($n = 1, 2, \ldots$) of nonlinear ordinary differential functions $H_n^{[m]}$, starting from $H_1^{[m]}$ now written differently (about their origin see Appendix)

$$H_1^{[m]} = (\ln u)'' + H_0^{[m]} \qquad H_0^{[m]} = u^m$$
$$H_2^{[m]} = (\ln u H_1^{[m]})'' + H_1^{[m]}$$
$$H_3^{[m]} = (\ln u H_1^{[m]} H_2^{[m]})'' + H_2^{[m]},\ \text{etc.}$$

For $n > 1$ they have the same structure for both m. (Hereafter, for readability, we omit the superscript m in all entities marked by it, if both m are applicable or if it is clear from context which of them in use). The general form of the n-th function is

$$H_n = (\ln y_n)'' + H_{n-1} \qquad y_n = u H_1 H_2 \ldots H_{n-1} = y_{n-1} H_{n-1} \qquad (7a, b)$$

where $y_1 = u$ for both m. These equalities lead to the three-term recurrence

$$H_n = (\ln H_{n-1})'' + 2 H_{n-1} - H_{n-2} \qquad (8)$$

Using $H_n$, we can derive expressions that contain (3) as particular case. Let $r_n = 1/y_n$, $l_n = \ln y_n$. Like in §1, the formula $r_n Q y_n = Q l_n + 3 l_n' l_n'' + l_n'^3$ gives $r_n Q y_n + y_n Q r_n = 6 l_n' l_n''$. This, with $l_n'' = H_n - H_{n-1}$ from (7a), yields

$$r_n U_n + y_n R_n = 0 \qquad U_n = Q y_n + 6 H_{n-1} y_n' \qquad R_n = Q r_n + 6 H_n r_n' \qquad (9)$$

Thus, each $U^{[m]}$ in (2) is represented by (9) as the first member of the infinite family of $U_n^{[m]}$.

Find the relation between $U_n$. Apply Q to $y_{n+1} = y_n H_n$, by (7b), and use (7a) to obtain

$$Q y_{n+1} = y_n Q H_n + H_n Q y_n + 3(y_n' H_n')' \qquad Q H_n = Q(l_n'' + H_{n-1})$$

Take $Q l_n$ from the above formula for $r_n Q y_n$ and express $Q y_i$ through $U_i$ given in (9) to get

$$U_{n+1} = y_n (r_n U_n)'' + H_n U_n + y_n P_n \qquad (10)$$
$$P_n \equiv Q H_{n-1} + 3(H_{n-1}^2 + l_n'^2 H_{n-1} - l_n' H_{n-1}')' \qquad (11)$$

Writing here $Q H_{n-1} = Q y_n r_{n-1}$ leads to $P_n(U_n)$

$$y_n P_n \equiv H_{n-1}(U_n - H_{n-1} U_{n-1}) \qquad (12)$$

With that, (10) becomes the three-term recurrence for $n > 1$ (as $U_0$ is undefined)

$$U_{n+1} = y_n (r_n U_n)'' + (H_n + H_{n-1}) U_n - H_{n-1}^2 U_{n-1} \qquad (13)$$

Whereas for $n = 1$, from (11) with $H_0^{[m]} = u^m$ it follows

$$P_1^{[1]} = U_1^{[1]} \qquad P_1^{[2]} = Q u^2 + 3(u^4 - u'^2)' = 2 u U_1^{[2]}$$

and thereby the recurrence reduces in this case to the relation between $U_2$ and $U_1$ only

$$m = 1:\ U_2 = u(r U_1)'' + H_1 U_1 + u U_1 \qquad m = 2:\ U_2 = u(r U_1)'' + H_1 U_1 + 2 u^2 U_1 \qquad (14a, b)$$

Next, like in §1, consider two equations (for each m)

$$H_n = 0 \qquad Qr_n = 0 \qquad (15)$$

that lead to $R_n = 0$, and hence to $U_n = -y_n^2 R_n = 0$, due to (9). We refer to them as $CE_n^{[m]}$ (coupled equations) and to their mutual solution as $u_n(x, t)$. Here again, $H_n = 0$ defines the functional form of this solution, $u_n(x)$, and $Qr_n = 0$ its t-dependance. In subsequent sections we show that $u_n(x, t)$ represents n moving bumps and thus is the n-soliton. For now, via the following statement, we prove (albeit not directly) that all $u_n(x, t)$ satisfy $U_1 = 0$.

**Statement 1.** For all n, if $CE_n$ hold, then $U_1 = 0$.
Proof. Let $CE_n$ (15) hold for a given $n > 1$, then $U_n = 0$ and $U_{n+1} = 0$ too, as $y_{n+1} = y_n H_n$. With that, the recurrence (13) leads to $U_{n-1} = 0$. Further, with $U_n$ and $U_{n-1}$ being zero, it leads to $U_{n-2} = 0$, and so on. The final triplet with the lowest indices: $U_3 = 0$, $U_2 = 0$ and $U_1$ (the same as for $n = 2$) gives the sought $U_1 = 0$. The case $n = 1$ follows directly from (9) as in §1.

**3.** In fact, $U_n$ (and paired with them $R_n$) in (9) are constructed by the use of functions $H_n^{[m]}$ and the operator $Q = \partial_t + \partial^3$. In this section we employ the same $H_n^{[m]}$ but other linear partial operators Qs to construct Us that admit the separation of variables described above and expressed by (15).
   We say a U admits SV, if it can be represented as the first member of an infinite set of $U_n$ that satisfy the three-term recurrence (13) (that is, form it without a remainder), and hence the Statement 1 holds. Such U is of Type-1, 2, if $m = 1, 2$ in $H_1^{[m]}$ (the lhs's of the KdV and mKdV are the examples of these two types).
   In order for such SV to work, an operator Q should have (i) constant coefficients, (ii) the time derivative(s) of the first order only, (iii) the orders of all its derivatives of the same parity. The last restriction comes from the necessity to satisfy the PDE of each $CE_n$, $Qr_n = 0$, where $r_n$, arising from $H_n = 0$, is a sum of $b_i \cosh(k_i x - \omega_i t)$, $b_i = $ const (§5). Below we use Qs that meet these conditions and call them *valid*.
   Each U we intend to obtain is supposed to have the form $U = Qu + N$, where N, nonlinear terms, is a polynomial in $u(x, t)$, its x- and t-derivatives and $\int (u^m)_t dx$.
   The construction stems from calculation of $r_n Q y_n \pm y_n Q r_n$ with +/– when derivatives in Q are odd/even. Such expression is a polynomial in derivatives of $l_n = \ln y_n$. Substituting $l_n'' = H_n - H_{n-1}$ or $(l_n)'_t = (h_n - h_{n-1})_t$, $h_n \equiv \int H_n dx$, into it provides $r_n U_n \pm y_n R_n = 0$, where $U_n = Q y_n + N_n(y_n, H_{n-1})$ and $R_n = Q r_n + M_n(r_n, H_n)$. The recurrence (13) has the same structure for any valid operator Q, as it always comes from $Q y_{n+1} = Q y_n H_n = H_n Q y_n + y_n Q(l_n'' + y_n r_{n-1}) + \ldots$ with $Q y_i$ and $Q r_i$ expressed through $U_i$; the term $Q l_n''$ yields $(r_n U_n)''$.

   We begin with two valid Qs containing again odd derivatives. Although the construction is based on the case $m = 1$ ($l'' = H_1 - u$), its results cover both m. From $r_n Q y_n + y_n Q r_n$ we obtain $S_n \equiv r_n U_n + y_n R_n = 0$. The crucial step is checking whether $U_2$ and $U_1$ from these $U_n$ are related by (14a). It comes from $Q y_2 = Q H_1 u = H_1 Q u + u Q(l'' + u) + \ldots$; the terms not containing $U_{1,2}$ should vanish. If they do not, we regroup the terms in $S_n$ trying first of all to meet (14a). Then, if (i) new $U_2(y_2, H_1)$ satisfies (14a), (ii) it reduces to $U_1$ under replacements $y_2 \to u$, $H_1 \to H_0$, (iii) $U_2$ and $R_1$ have the same structure, then construction is on the right way. (In the first line of each example we place the formula by which $r_n Q y_n$ can be expressed through $l_n$).

- $Q = -\partial_t - \partial + \partial_t \partial^2$ $\qquad\qquad r u_t'' = l_t''' + 2 l' l_t' + l_t(l'' + l'^2)$

From $r_n Q y_n + y_n Q r_n = 2(l_n)_t l_n'' + 4 l_n'(l_n)_t'$ we get $S_n = 0$, where

$$U_n = Q y_n + 2(y_n)_t H_{n-1} + 4 y_n'(h_{n-1})_t \qquad\qquad R_n = Q r_n + 2(r_n)_t H_n + 4 r_n'(h_n)_t$$

Calculation shows that such $U_{1,2}$ do not satisfy (14a). To find a fitting, we take the trial
$$U(a, b) = Qu + auu_t + bu'\int u_t dx$$
with the same terms as in $U_1$ but arbitrary constant coefficients. It turns out they should be $a = 4$, $b = 2$ to eliminate nonpolynomial terms in $U_2$ and make it reducible to $U$. In view of this, we rearrange the terms in $S_n = 0$ accordingly
$$S_n \equiv r_n[Qy_n + 4(y_n)_t H_{n-1} + 2y_n'(h_{n-1})_t] + y_n[Qr_n + 4(r_n)_t H_n + 2r_n'(h_n)_t] - 2Z_n = 0$$
$$Z_n = (l_n)_t H_{n-1} - l_n'(h_{n-1})_t + l_n'(h_n)_t - (l_n)_t H_n$$
Differentiation of $Z_n$ helps reveal two parts contributing to $U_n$ and $R_n$
$$Z_n' = A_U + A_R \qquad A_U = (l_n)_t H_{n-1}' - l_n'(H_{n-1})_t \qquad A_R = l_n'(H_n)_t - (l_n)_t H_n'$$
We include $\int A_{U,R} dx$ into the groups in $S_n$ with $Qy_n$ and $Qr_n$, respectively, to finally get new
$$U_n = Qy_n + 4(y_n)_t H_{n-1} + 2y_n'(h_{n-1})_t + 2y_n \int [l_n'(H_{n-1})_t - (l_n)_t H_{n-1}']dx \qquad (16)$$
$$R_n = Qr_n + 4(r_n)_t H_n + 2r_n'(h_n)_t - 2r_n \int [l_n'(H_n)_t - (l_n)_t H_n']dx$$
Such $U_n$ do satisfy the recurrence (13). Hence, their first member $U(4, 2)$ (its integrand in (16) is zero) admits SV, Type-1. (Although $U_1$ and $R_1$ have different structures, unlike the (m)KdV, this fact by no means contradicts the general scheme). The equation $U(4, 2) = 0$ is known as the Shallow Water Wave (SWW) equation [3].

When $m = 2$ ($H_0 = u^2$), the recurrence (13) for these $U_n$ holds as well, with $U_2$ and $U_1$ being related by (14b). In this case, the first member of (16) is of Type-2
$$Qu + 4u^2 u_t + 2u'\int (u^2)_t dx$$
from the so called modified SWW equation [4].

- $Q = \partial_t + \partial^5 \qquad\qquad ru^v = l^v + 5l'l^{iv} + 10(l'''l'^2 + l''l'^3 + l'l''') + 15l'l''^2 + l'^5$
$$r_n Qy_n + y_n Qr_n = 10l_n'l_n^{iv} + 20l_n''(l_n'^3 + l_n''') \equiv G_n$$
For $l_n'^3$ and $l_n'^2$ we use the writing
$$l_n'^3 = r_n y_n''' - l_n''' - 3l_n'l_n'' = - y_n r_n''' - l_n''' + 3l_n'l_n''$$
$$l_n'^2 = r_n y_n'' - l_n'' = y_n r_n'' + l_n''$$
In $G_1$, we write $20l'^3$ as the sum $(\gamma + c)l'^3$, $\gamma + c = 20$, with the terms $\gamma l'^3 = \gamma(ru''' - l''' - 3l'l'')$ and $cl'^3 = cl'l'^2 = cl'(ru'' - l'')$ to get $20l'^3 = (20 - \gamma)l''' + \gamma ru''' - (3\gamma + c)l'l'' + cl'ru''$. Then
$$G_1 = 10l'l^{iv} + l''[(20 - \gamma)l''' + \gamma ru''' - (2\gamma + 20)l'l'' + (20 - \gamma)l'ru'']$$
Setting here $l'' = - u$ ($H_1 = 0$) leads to the trial
$$U(\gamma) = Qu + N(\gamma), \qquad N(\gamma) = \gamma uu''' + (30 - \gamma)u'u'' + 3\gamma u^2 u' = [\gamma(uu'' + u^3) + (15 - \gamma)u'^2]'$$
It turns out, only if $\gamma = 10$, then $U_2$ is reducible to $U_1 = U(10)$ and $U_{1,2}$ satisfy (14a). In such a case, $S_1 = 0$ involves
$$U_1 = Qu + 10uu''' + 30u^2 u' + 10(u'^2)'$$
$$R_1 = Qr + 10H_1 r''' + 30H_1^2 r' + 10(H_1'r')' - 10r'^2(H_1/r)'$$
$S_2 = 0$ involves
$$U_2 = Qy_2 + 10H_1 y_2''' + 30H_1^2 y_2' + 10(H_1'y_2')' - 10y_2'^2(H_1/y_2)' = u(rU_1)'' + H_1 U_1 + uU_1$$
$$R_2 = Qr_2 + 10H_2 r_2''' + 30H_2^2 r_2' + 10(H_2'r_2')' - 10r_2'^2(H_2/r_2)'$$
and $S_n = 0$, replicating the structure of $S_2 = 0$, involves
$$U_n = Qy_n + 10H_{n-1} y_n''' + 30H_{n-1}^2 y_n' + 10(H_{n-1}'y_n')' - 10y_n'^2(H_{n-1}/y_n)' \qquad (17)$$
$$R_n = Qr_n + 10H_n r_n''' + 30H_n^2 r_n' + 10(H_n'r_n')' - 10r_n'^2(H_n/r_n)'$$
These $U_n$ satisfy the recurrence (13) (checking this is a quite laborious task). So, their first member $U(10)$ admits SV, Type-1. This is the lhs of the second equation of the KdV hierarchy.

When $m = 2$, the first member of (17) is of Type-2
$$Qu + 10u^2 u''' + 30u^4 u' + 40uu'u'' + 10u'^3$$
from the second equation of the mKdV hierarchy.

- $Q = \partial_t \partial + c$                          $ru'_t = l'_t + l'l_t$

Here the valid Q contains the even derivative and a constant (multiplication operator).
From $r_n Q y_n - y_n Q r_n = 2(l_n)'_t = 2(h_n - h_{n-1})_t$ follows $r_n U_n - y_n R_n = 0$, where

       $U_n = Q y_n + 2 y_n (h_{n-1})_t$          $R_n = Q r_n + 2 r_n (h_n)_t$

Calculation of $Q y_{n+1}$ gives (10) with

       $P_n = - (H_{n-1})'_t + (l_n)_t H_{n-1}' + l_n'(H_{n-1})_t + 2 H_{n-1}(l_n)'_t$

Its form $P_n(U_n)$ is given by (12). So, the recurrence (13) holds for both m. However, the needed relation between $U_2$ and $U_1$ in (14) holds only for m = 2 since $P_1^{[2]} = 0$, but $P_1^{[1]} = u'_t$. Thus, only

       $Qu + 2u\int (u^2)_t dx$

of Type-2 admits SV. (Moreover, for all Qs with even derivatives including a constant, the resulting U cannot be of Type-1 primarily because the PDEs $Qr_n^{[1]} = 0$ cannot be satisfied as each $r_n^{[1]}$ is a sum of $b_i \cosh(k_i x - \omega_i t)$ plus a constant, see (24)).

    Such U appears in the simplest version of the Self-Induced Transparency problem [2]. Remark, it can be transformed into the Sin-Gordon equation $g'_t = \pm c \sin g$, $g' = \pm 2u$, to which this SV cannot be applied directly.

**4.** To find $u_n(x)$, the solution of $H_n = 0$, we employ the following approach. We introduce the functions $\chi_n$ such that $H_n = (\ln \chi_n)''$. Then (7) and (8) become

       $\chi_n = y_n \chi_{n-1}$     $\chi_n = H_{n-1} \chi_{n-1}^2 / \chi_{n-2} = D\chi_{n-1}/\chi_{n-2}$     $Df(x) \equiv ff'' - f'^2 = f^2 (\ln f)''$     (18)

Therewith the sought function acquires the twofold representation

       $u \equiv y_1 = \chi_1 / \chi_0$         $u^m = H_0 = (\ln \chi_0)'' = D\chi_0 / \chi_0^2$                       (19)

Remark, to get (18), we assumed that f = 1 in $(\ln f)'' = 0$. Likewise, we consider identical the equalities $H_n = 0$, $\int H_n dx = 0$ and $\chi_n = 1$ (the use of general $\chi_n = e^{qx+c}$ modifies intermediate formulas obtained with $\chi_n = 1$, but not the ultimate form of $u_n(x)$).

    When $\chi_n = 1$ ($H_n = 0$), then $\chi_1$ and $\chi_0$, forming u, become the differential functions of $\chi_{n-1}$. Indeed, when $\chi_n = 1$ in $\chi_{n-2} = D\chi_{n-1}/\chi_n$, given by (18), then all $\chi_{n-j}$ become functions of $\chi_{n-1}$ only. The first few of them are

       $\chi_{n-2} = D\chi_{n-1}$                                                                                  (20)

       $\chi_{n-3} = D\chi_{n-2}/\chi_{n-1} = D^2\chi_{n-1}/\chi_{n-1} \equiv \Phi(\chi_{n-1})$

       $\chi_{n-4} = D\chi_{n-3}/\chi_{n-2} = D\Phi(\chi_{n-1})/D\chi_{n-1}$

For each n in $\chi_n = 1$ (and both m), they define $\chi_{1,0}(\chi_{n-1})$

       n = 2:   $\chi_0 = D\chi_1$                                                                          (21)

       n = 3:   $\chi_1 = D\chi_2$, $\chi_0 = \Phi(\chi_2)$

       n = 4:   $\chi_1 = \Phi(\chi_3)$, $\chi_0 = D\Phi(\chi_3)/D\chi_3$

and thereby $u_n(x) = \chi_1/\chi_0$ with such $\chi_{1,0}(\chi_{n-1})$.

    Each $u_n(x)$ can be written as a simple ratio $u_n(x) = \Pi_1/\Pi_0$ of two differential polynomials of $\chi_{n-1}$. For the first three such a form is straightforward

       $u_1 = 1/\chi_0$                 $u_2 = \chi_1/D\chi_1$           $u_3 = D\chi_2/\Phi(\chi_2) = \chi_2 D\chi_2/D^2\chi_2$      (22)

Further the formula $D(a/b) = (b^2 Da - a^2 Db)/b^4$ should be used, like in

       $u_4 = \Phi/(D\Phi/Df) = (D^2 f)(Df)f^3/W$

where $f \equiv \chi_3$, $\Phi = \Phi(f)$ and $W \equiv f^2 D^3 f - (D^2 f)^2 Df$ is from $D\Phi \equiv D(D^2 f/f) = W/f^4$.

    As an alternative, one can use the fact that $\Phi(f)$ is differential polynomial

       $\Phi(f) \equiv D^2 f/f = f^{iv} Df - f'''(Df)' + f'' Df'$                $D^2 f = D(Df) = (Df)(Df)'' - (Df)'^2$

(There are good grounds to conjecture that all ratios in (20) are in fact polynomials of $\chi_{n-1}$).

Still unknown $\chi_{n-1}$ is to be found by solving an ODE that arises as follows. Combining the two forms of u in (19) yields $D\chi_0/\chi_0^2 = (\chi_1/\chi_0)^m$ and thereby the relations between $\chi_0$ and $\chi_1$

$\quad$ m = 1: $\ D\chi_0 = \chi_1\chi_0$ $\qquad\qquad\qquad$ m = 2: $\ D\chi_0 = \chi_1^2$

These together with (21), resulting from $H_n = 0$, lead to the sought nonlinear ODEs for $\chi_{n-1}$ (the left column always corresponds to m = 1)

$\quad \chi_1 = 1:\ \ D\chi_0 = \chi_0 \qquad\qquad\qquad\qquad D\chi_0 = 1$
$\quad \chi_2 = 1:\ \ D^2\chi_1 = \chi_1 D\chi_1 \qquad\qquad\quad\ D^2\chi_1 = \chi_1^2$ $\qquad\qquad$ (23)

The next pair, $D\Phi(\chi_2) = \Phi D\chi_2$ and $D\Phi(\chi_2) = (D\chi_2)^2$ written in one tier form is, $f = \chi_2$

$\quad \chi_3 = 1:\ \ W = f^3(Df)D^2f \qquad\qquad\qquad W = f^4(Df)^2$

We call them NEs (nonlinear equations), and denote the one corresponding to $\chi_n = 1$ as $NE_n^{[m]}$. When $\chi_n = 1$, then $r_n = \chi_{n-1}$, by (18), so the PDE in (15) becomes $Q\chi_{n-1} = 0$. For each n, this PDE together with $NE_n$ (both for the same $\chi_{n-1}$) comprise an alternative form of $CE_n$ (15).

As it is shown in §5. 6, the solution of $NE_n$, $\chi_{n-1}(x)$, that could give rise to the n-soliton is a finite sum of $b_i\cosh(k_i x + x_i)$. Setting in it $x_i = -\omega_i t$ eventually defines the t-dependance in $u_n(x, t)$ given by (22). The only contribution of the linear PDE $Q\chi_{n-1}(x, t) = 0$ to solving $CE_n$ is to determine a (dispersion) function $\Omega(\kappa)$ in $\omega_i = \Omega(k_i)$ via $Q\cosh(\kappa x - \Omega t) = 0$. (For example, $Q = \partial_t + \partial^3$ and $Q = -\partial_t - \partial + \partial_t\partial^2$ yield $\Omega(\kappa) = \kappa^3$ and $\Omega(\kappa) = \kappa/(1 - \kappa^2)$). With this done, the above sum represents the solution $\chi_{n-1}(x, t)$ of $CE_n$. In view of such an auxiliary role of the PDE, the main emphasis will be given to finding $u_n(x)$.

**5.** The procedure of finding $\chi_{n-1}(x)$ and then $u_n(x)$ is the following. Seek for a solution of each $NE_n$ in (23) in the form (justified in §6, 7)

$\quad \chi_{n-1}(x) = \Sigma b_i \cosh \eta_i + b_0 \qquad\qquad \eta_i = k_i x + x_i \qquad\qquad\qquad (24)$

The sum contains n terms for both m, and $b_0 = 0$ for m = 2. Substitute it into $NE_n$ and equate coefficients of like terms (cosh's and sinh's with arguments $\eta_i$, $\eta_i \pm \eta_j$, $\eta_i \pm \eta_j \pm \eta_l$ etc., the variety of such combinations increasing with n). This defines $b_i(k_i)$ and thereby $\chi_{n-1}(x)$ that we call the general solution of $NE_n$ ($k_i$ are arbitrary). It gives rise, via (22), to general $u_n(x) = \chi_1/\chi_0 = \Pi_1/\Pi_0$.

Here we find the general solutions of $NE_{1,2}^{[m]}$. We use some formulas with the operator D
$\quad Dfg = f^2 Dg + g^2 Df \qquad\qquad D(f/g) = (g^2 Df - f^2 Dg)/g^4$
$\quad D(f + g) = Df + Dg + fg'' + gf'' - 2f'g' \qquad\qquad\qquad\qquad\qquad\qquad (25a)$
$\quad D(a \exp kx + b \exp qx) = ab(k - q)^2 \exp(k + q)x \qquad D(\exp kx) = 0$
$\quad D(a \cosh kx) = a^2 k^2 \qquad\qquad\qquad\qquad\qquad\qquad\qquad\qquad\qquad (25b)$
$\quad Df^s = sf^{2(s-1)}Df \qquad\qquad D(\cosh kx)^s = sk^2(\cosh kx)^{2(s-1)} \qquad\qquad (25c)$

<u>n = 1, m = 2</u>. From $NE_1^{[2]}$, $D\chi_0 = 1$, by (25b): $\chi_0 = \pm (1/k)\cosh(kx + x_2)$. Then, with $\chi_1 = 1$, we have $u_1^{[2]} = 1/\chi_0 = \pm k \operatorname{sech}(kx + x_2)$, cf. (1).

<u>n = 1, m = 1</u>. Substitute the ansatz $\chi_0 = b_1\cosh(kx + x_1) + b_0$ into $NE_1^{[1]}$, $D\chi_0 = \chi_0$, and use (25a) to find $b_0 = 1/k^2$, $b_1^2 = 1/k^4$. So, $\chi_0 = (1/k^2)[1 \pm \cosh(kx + x_1)]$. With plus sign and 2k instead of k, the function $u_1^{[1]} = 1/\chi_0$ has the form (1).

<u>n = 2, m = 2</u>. Seek for the solution of $NE_2^{[2]}$
$\quad D^2\chi_1 = \chi_1^2 \qquad\qquad\qquad\qquad\qquad\qquad\qquad\qquad\qquad\qquad\qquad (26)$
in the form $\chi_1 = b_1\cosh \eta_1 + b_2\cosh \eta_2$. Then, see (21),

$$\chi_0 = D\chi_1 = (b_1 b_2/2)[\Delta^2 \cosh(\eta_1 + \eta_2) + \Sigma^2 \cosh(\eta_1 - \eta_2)] + C$$

where $\Delta \equiv k_1 - k_2$, $\Sigma \equiv k_1 + k_2$, $C \equiv (b_1 k_1)^2 + (b_2 k_2)^2$. Substitute these $\chi_1$ and $D\chi_1$ into (26) to get $(b_1 k_1)^2 = (b_2 k_2)^2 = 1/(\Sigma\Delta)^2$. As a result, $u_2^{[2]}(x) = \chi_1/D\chi_1 = 2\Sigma\Delta(F_1/F_0)$, where

$$F_1 = k_2 \cosh(k_1 x + x_1) + k_1 \cosh(k_2 x + x_2) \tag{27}$$
$$F_0 = \Delta^2 \cosh(\Sigma x + x_1 + x_2) + \Sigma^2 \cosh(\Delta x + x_1 - x_2) + 4 k_1 k_2$$

<u>$n = 2$, $m = 1$</u>. Seek for the solution of $NE_2^{[1]}$

$$D^2 \chi_1 = \chi_1 D\chi_1 \tag{28}$$

in the form $\chi_1 = b_1 \cosh \eta_1 + b_2 \cosh \eta_2 + b_0$. Then $\chi_0 = D\chi_1$, found in the same way and notation as for $m = 2$, is

$$\chi_0 = (b_1 b_2/2)[\Delta^2 \cosh(\eta_1 + \eta_2) + \Sigma^2 \cosh(\eta_1 - \eta_2)] + b_0 (b_1 k_1^2 \cosh \eta_1 + b_2 k_2^2 \cosh \eta_2) + C$$

where $(b_1 k_1 \Sigma\Delta)^2 = b_0 k_2^2$, $(b_2 k_2 \Sigma\Delta)^2 = b_0 k_1^2$, $(b_0 k_1^2 k_2^2)^2 = 1$. And $u_2^{[1]}(x) = \chi_1/D\chi_1 = 2\Sigma\Delta(G_1/G_0)$

$$G_1 = k_2^2 \cosh(k_1 x + x_1) + k_1^2 \cosh(k_2 x + x_2) + \Sigma\Delta \tag{29}$$
$$G_0 = \Delta^2 \cosh(\Sigma x + x_1 + x_2) + \Sigma^2 \cosh(\Delta x + x_1 - x_2) + 2\Sigma\Delta[\cosh(k_1 x + x_1) + \cosh(k_2 x + x_2)] + 2(k_1^2 + k_2^2)$$

Remark, such signs of coefficients in both $u_2^{[m]}(x)$ correspond to positive square roots in the relations defining $b_i$.

The resulting $u_2^{[1, 2]}(x, t)$, with $x_i = -\Omega(k_i)t$ where $\Omega(k_i)$ is found from $Q\chi_1 = 0$, are the two-solitons in their general form. The expressions similar or equivalent to them obtained by various methods are available in literature, see e.g. [2].

**6.** Each $NE_n$ has the special solution – such that, still being a finite sum of $b_i \cosh(k_i x + x_i)$, becomes $\cosh^\lambda kx$ when $x_i = 0$, $\lambda(n)$ is a positive integer. (In this section we deal only with such special solutions – without introducing a specific notation for them).

First we derive such solution of $NE_2^{[2]}$ (26) as a particular case of its general (27). With $k_1 = 3k$ and $k_2 = k$, $F_{1,0}$ become

$$F_1(x, x_i)/k_2 = \cosh(3kx + x_1) + 3\cosh(kx + x_2), \qquad F_1(x, 0) \propto \cosh^3 kx \tag{30}$$
$$F_0(x, x_i)/\Delta^2 = \cosh[4kx + (x_1 + x_2)] + 4\cosh[2kx + (x_1 - x_2)] + 3, \quad F_0(x, 0) \propto \cosh^4 kx$$

It is easy to check via (25c) that $f = \cosh^3 kx$ satisfies (26): $Df \propto \cosh^4 kx \propto F_0(x, 0)$, and finally, $D^2 f \propto \cosh^6 kx = f^2$.

Now we obtain the same result by the procedure that can be applied to each NE. Seek for a solution of (26) in the form $\cosh^\lambda kx$. Use (25c) to get the equation $2[2(\lambda - 1) - 1] = 2\lambda$, giving $\lambda = 3$. Write $4\cosh^3 kx = \cosh 3kx + 3\cosh kx$, and then add $x_{1,2}$ to arguments of cosh's. Although the resulting sum is not a powered cosh anymore, it still satisfies NE, as it is nothing else then $F_1(x, x_i)$ in (30), which does. Thus formed sum is the desired special solution. Acting by $D$ on it yields $F_0(x, x_i)$ in (30).

Likewise, $\lambda = 4$ found from $NE_2^{[1]}$ (28) gives rise to the special $G_{1,0}$ corresponding to general ones in (29) with $k_1 = 4k$ and $k_2 = 2k$

$$G_1(x, x_i)/k_2^2 = \cosh(4kx + x_1) + 4\cosh(2kx + x_2) + 3, \qquad G_1(x, 0) \propto \cosh^4 kx \tag{31}$$
$$G_0(x, x_i)/\Delta^2 = \cosh[6kx + (x_1 + x_2)] + 9\cosh[2kx + (x_1 - x_2)] + 6\cosh(4kx + x_1) + 6\cosh(2kx + x_2) + 10, \qquad G_0(x, 0) \propto \cosh^6 kx$$

The power $\lambda(n)$ can be found in a general manner. For this, substitute $u^{[1]} = a_1 \text{sech}^2 kx$ into $H_n^{[1]}$, and $u^{[2]} = a_2 \text{sech} \, kx$ into $H_n^{[2]}$ to get

$$H_n^{[1]} = [a_1 - n(n+1)k^2] \, \text{sech}^2 kx \qquad H_n^{[2]} = [a_2^2 - n^2 k^2] \, \text{sech}^2 kx$$

So, $H_n^{[1, 2]} = 0$ when $a_1 = n(n + 1)k^2$ and $a_2 = \pm nk$. With such $a_m$

$$H_{n-1}^{[1]} = 2nk^2\text{sech}^2 kx \qquad\qquad H_{n-1}^{[2]} = (2n-1)k^2\text{sech}^2 kx$$

Setting $\chi_{n-1} \propto \cosh^\lambda kx$ in each $H_{n-1} = (\ln \chi_{n-1})''$ yields the sought

$$\lambda^{[1]} = 2n \qquad\qquad \lambda^{[2]} = 2n-1 \tag{32}$$

The sum equivalent to the powered cosh comes from the expansion of $(e^{kx} + e^{-kx})^\lambda$

$$\chi_{n-1}(x) \propto 2^{\lambda-1}\cosh^\lambda kx = \Sigma b_i \cosh(\gamma_i kx) + b_0 \tag{33}$$

$b_i$ are the binomial coefficients: $b_1 = 1$, $b_2 = \lambda$, $b_3 = \lambda(\lambda-1)/2$, etc.; $\gamma_1 = \lambda$, $\gamma_2 = \lambda - 2$, $\gamma_3 = \lambda - 4$ etc., that is, $\gamma_i$ are even (odd) numbers when $m = 1$ ($m = 2$), according to (32). The number of cosh's in such sum is n for both m; $b_0^{[2]} = 0$ and $b_0^{[1]}$ is the coefficient of $e^0/2$ in the expansion.

The sum (33) modified by appending $x_i$ to arguments $\gamma_i kx$ represents the special solution of $NE_n$ and, clearly, is a particular form of a yet unknown general one; hence the ansatz (24). This modified sum, with the t-dependance introduced via $x_i = -\omega_i t$ with $\omega_i = \Omega(\gamma_i k)$ found from $Q\chi_{n-1} = 0$, represents the special solution of $CE_n$, b = const,

$$b\chi_{n-1}(x, t) = \Sigma b_i \cosh[\gamma_i kx - \Omega(\gamma_i k)t] + b_0 \tag{34}$$

Finally, by formulas (22) it yields the special $u_n(x, t) = \chi_1/\chi_0 = \Pi_1/\Pi_0$.

At $t = 0$, $\Pi_{1,0}$ turn into powered cosh's: $\Pi_1(x, 0) \propto \cosh^\alpha kx$ and $\Pi_0(x, 0) \propto \cosh^\beta kx$. Indeed, like in (30) and (31), the action of operators in (22) on the sum (34) and then setting $t = 0$ provides the same result as their action on $\cosh^\lambda kx$ given by formula (25c). They form $u_n^{[m]}(x, 0)$, the solutions of $H_n^{[m]} = 0$ with $a_m$ found above (let $a_2 > 0$)

$$u_n^{[1]}(x, 0) = n(n+1)k^2\text{sech}^2 kx \qquad\qquad u_n^{[2]}(x, 0) = nk\,\text{sech}\, kx \tag{35}$$

(the reflectionless potentials, in terms of scattering problems). Therefore, $\beta - \alpha = 2, 1$ for $m = 1, 2$. At $t = 0$, the special $\chi_{1,0}$ are also powered cosh's. The power p(n) in $\chi_0(x, 0) \propto \cosh^p kx$ follows from the representation $u^m = (\ln \chi_0)''$

$$p^{[1]} = n(n+1) \qquad\qquad p^{[2]} = n^2$$

For each particular n, $\beta$ and p can be related. For example, $u_3$ has $\Pi_1 = \chi_2 D\chi_2$ and $\Pi_0 = D^2\chi_2$. By (25c): $\alpha = \lambda + 2(\lambda - 1)$ and $\beta = 2[2(\lambda - 1) - 1]$. From (32): $\lambda^{[1, 2]} = 6, 5$, so $\alpha^{[1, 2]} = 16, 13$ and $\beta^{[1, 2]} = 18, 14$. On the other hand, since $\Pi_0 = \chi_0\chi_2$, then $\beta = p + \lambda$. This, with $p^{[1, 2]}(n = 3) = 12, 9$, provides the same $\beta^{[1, 2]}$.

In proving that the special $u_n^{[m]}(x, t)$ indeed represent the n-solitons we rely on their graphs plotted for n = 2, 3, 4 and both m, with $\Omega(\kappa) = \kappa^3$. They all have similar commonly known pattern (like in picture). Namely, the graph of $u_n^{[m]}(x, t > 0)$ depicts n bumps, that develop from $u_n^{[m]}(x, 0)$ in (35), move with different velocities, the higher the faster, to the right (due to the chosen sign in the t-dependance) and become well separated at large t. Apparently, they are multisolitons. (The x-derivatives of $\chi_{n-1}$ have been calculated manually, the computer program only performed plotting of $u_n(x, t) = \chi_1/\chi_0$).

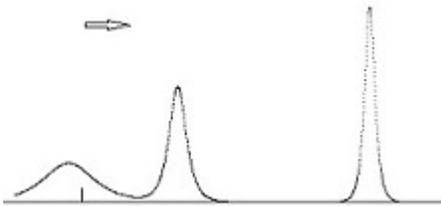

The picture depicts the special 3-soliton of mKdV:

$$u_3 = \chi_2 D\chi_2/D^2\chi_2 \qquad\qquad \chi_2(x, 0) \propto \cosh^5 x$$

$\chi_2(x, t) \propto \cosh(5x - 5^3 t) + 5\cosh(3x - 3^3 t) + 10\cosh(x - t)$, $\quad t = 0.5$

The general solutions of $CE_n$'s are richer, of course, than special ones. The graphs of $u_2^{[m]}(x, t)$ from (27) and (29) show that varying $k_1/k_2$ leads, in some ranges, to considerable changes in their evolution in time.

**7.** Here, proceeding from $H_n = 0$ with $n = 1, 2$ and both m, we derive the linear ODEs for $\chi_{n-1}(x)$, for the same functions that are involved in $NE_n$. We call them LEs (linear equations) and denote as $LE_n$ the one resulting from $H_n = 0$. Their coefficients are constant, all derivatives are even and the order is 2n (that of the leading derivative in the $NE_n$). They are homogeneous for m = 2 and nonhomogeneous for m = 1. (All this is also true for $LE_3^{[m]}$ that we derived too, in the slightly different way, but do not present here to save space because and this case does not give a hint for generalization on n). Such reduction of nonlinear ODEs to linear being remarkable, we present our calculations in more details.

Notations: $h_n \equiv \int H_n dx = (\ln \chi_n)'$, $\varphi_n \equiv \chi_n''/\chi_n = H_n + h_n^2$; the equalities $H_n = 0$, $h_n = 0$ and $\chi_n = 1$ are identical as before.

We integrate (8) to get, for n > 1
$$h_n = (\ln H_{n-1})' + 2h_{n-1} - h_{n-2} \tag{36}$$
and then multiply this by $H_{n-1}$
$$H_{n-1}h_n = \varphi_{n-1}' - H_{n-1}h_{n-2} \tag{37}$$
The case n = 1 goes separately – by using $H_0^{[m]} = (\ln \chi_0)'' = u^m$
$$m = 2: \quad 2H_0h_1 \equiv 2u^2(l' + h_0) = (H_0 + h_0^2)' = \varphi_0' \tag{38a}$$
$$m = 1: \quad 2H_0h_1 \equiv 2u(l' + h_0) = (2u + h_0^2)' = (\varphi_0 + u)' \tag{38b}$$
It is these two distinct forms that make the prospective LEs different for m = 2 and 1.

Set $h_n = 0$, then differentiate (37) and take $H_{n-1}'$ from (36) to get the equation (for both m)
$$\chi_{n-1}^{iv}/\chi_{n-1} - \varphi_{n-1}^2 - H_{n-1}\varphi_{n-2} = 0 \tag{39}$$

THE CASE m = 2

$\underline{h_1 = 0}$. $LE_1^{[2]}$ follows immediately from integration of (38a), $\varphi_0' = 0$, (cf. (6))
$$\chi_0'' - c_0\chi_0 = 0$$
$\underline{h_2 = 0}$. Equation (39) reads
$$\chi_1^{iv}/\chi_1 - \varphi_1^2 - H_1\varphi_0 = 0$$
We differentiate it, then use formulas $\varphi_0' = 2H_0h_1$, $0 = (\ln H_1)' + 2h_1 - h_0$ and $H_1h_0 = \varphi_1'$ from (38a), (36) and (37) to obtain
$$(\chi_1^{iv}/\chi_1)' - G\varphi_1' = 0 \qquad G = 2(\varphi_1 - h_1h_0) + \varphi_0 \tag{40}$$
where $G' = 0$ due to (37) and (38a). So, $G = c_2 = $ const and then integration of (40) with $b_2 = $ const yields $LE_2^{[2]}$
$$\chi_1^{iv} - c_2\chi_1'' - b_2\chi_1 = 0$$

THE CASE m = 1

Here LEs differ from those above only in that they involve an additional constant (namely + 1 or – 1) originated from the term *u* in (38b).

$\underline{h_1 = 0}$. From (38b), $(\varphi_0 + u)' = 0$, with $u = 1/\chi_0$ (as $\chi_1 = 1$), we have $LE_1^{[1]}$ (cf. (6))
$$\chi_0'' - c_0\chi_0 = -1$$
$\underline{h_2 = 0}$. The same calculations as for m = 2 but with the use of (38b) instead of (38a) give
$$(\chi_1^{iv}/\chi_1)' - G\varphi_1' + u'H_1 = 0$$
with the same G as in (40). But here $G' = -u'$, so $G = c_1 - u$. As a result, the equation contains the terms $u\varphi_1' + u'H_1 = uH_1(h_0 + l') = y_2h_1 = -(1/\chi_1)'$, as $y_2 = 1/\chi_1$ by (18). Thus
$$(\chi_1^{iv}/\chi_1)' - c_1\varphi_1' - (1/\chi_1)' = 0$$
Integration of this with $b_1 = $ const yields $LE_2^{[1]}$
$$\chi_1^{iv} - c_1\chi_1'' - b_1\chi_1 = 1$$

The above LEs can be equally derived from the equation following from (18) with $\chi_n = 1$

$$D\chi_{n-1} = \chi_{n-2} \quad (41)$$

For this, use formulas $(Df)' = ff''' - f'f'' = f^2(f''/f)'$, $(Df)'' = ff^{iv} - f''^2$.

Applying $(1/\chi_{n-1}^2)\partial$ to (41) yields the lhs $(D\chi_{n-1})'/\chi_{n-1}^2 = \varphi_{n-1}'$ and the rhs $\chi_{n-2}'/\chi_{n-1}^2 = h_{n-2}\chi_{n-2}/\chi_{n-1}^2$ and thus (37) with $h_n = 0$, as $\chi_{n-2}/\chi_{n-1}^2 = H_{n-1}$. Applying $(1/\chi_{n-1}^2)\partial^2$ yields $\chi_{n-1}^{iv}/\chi_{n-1} - \varphi_{n-1}^2$ and $\chi_{n-2}''/\chi_{n-1}^2 = \varphi_{n-2}\chi_{n-2}/\chi_{n-1}^2$ and thus (39).

Note, the ansatz (24) is nothing else than a solution to the above LEs, which is another (see §6) justification of its form.

**Appendix**

The functions $H_n^{[m]}$ are related to the Inverse Scattering Transform (IST) based on the Schrödinger equation (m = 1, IST-1) and the Dirac system (m = 2, IST-2). They appear in the process of solving the linear PDE (A3) with m = 1, 2 via step-by-step procedure called the Laplace cascade method (LCM) [5]. The PDE arises in the two Goursat problems [2]: for the kernel $K(x, \tau)$ of the integral equation of the IST-1; $u = u(x)$ is a real-valued function, scattering potential,

$$(\partial_x^2 - \partial_\tau^2)K + uK = 0, \qquad 2[(\partial_x + \partial_\tau)K(x, \tau)]_{x=\tau} = u(x)$$

and the kernels $K_{1,2}(x, \tau)$ from the IST-2

$$(\partial_x - \partial_\tau)K_1 = uK_2 \qquad (\partial_x + \partial_\tau)K_2 = -uK_1, \qquad 2K_1(x, x) = -u(x)$$

In terms of $y = (x + \tau)/2$ and $z = (x - \tau)/2$ they become, $u = u(y + z)$

$$K_{zy} + uK = 0, \qquad 2K_y(y, z)|_{z=0} = u(y) \qquad \text{(A1a, b)}$$
$$K_{1z} = uK_2 \qquad K_{2y} = -uK_1, \qquad 2K_1(y, 0) = -u(y) \qquad \text{(A2a, b, c)}$$

Applying $u\partial_y(1/u)$ to the equation (A1a) [to (A2a)] gives the following PDE with m =1 for $f \equiv K_y$ [with m = 2 for $f \equiv K_1$]

$$f_{zy} - (\ln u)_y f_z + u^m f = [f_y - (\ln u)_y f]_z + H_1^{[m]} f = 0, \qquad H_1^{[m]} \equiv (\ln u)_{yz} + u^m \qquad \text{(A3)}$$

Its second form represents the first step of the LCM (for such PDE). Further, applying $H_1\partial_y(1/H_1)$ to (A3) yields the similar equation for $\varphi \equiv f_y - (\ln u)_y f$, for both m,

$$\varphi_{zy} - (\ln uH_1)_y \varphi_z + H_1\varphi = [\varphi_y - (\ln uH_1)_y \varphi]_z + H_2\varphi = 0, \qquad H_2 \equiv (\ln uH_1)_{yz} + H_1 \quad \text{(A4)}$$

This is the second step of the LCM. The next step, via applying $H_2\partial_y(1/H_2)$ to (A4), provides $H_3$ as the coefficient of the last term in an equation of the same structure as (A3) and (A4); and so on.

E-mail: bryssev@gmail.com    17 May 2021